\documentclass[notitlepage,aps,prl,reprint,twocolumn,longbibliography,superscriptaddress]{revtex4-1}
\usepackage{graphicx}
\usepackage{amsmath}
\usepackage{amssymb}
\usepackage{comment}
\usepackage[colorlinks, allcolors=blue]{hyperref}
\usepackage[all]{hypcap}
\usepackage[mathlines]{lineno}
\usepackage{braket}
\usepackage{wrapfig}
\usepackage{lipsum}
\usepackage{ulem}

\newcommand{\pref}[2]{\hyperref[#1]{\ref{#1}(#2)}}
\newcommand{\preff}[2]{\hyperref[#1]{\ref{#1 b}#2}}
\newcommand{\eqpref}[1]{\hyperref[#1]{(\ref{#1})}}

\newcommand{\squig}{{\raise.17ex\hbox{$\scriptstyle\sim$}}}

\begin{document}
\title{Measuring the Adiabatic Non-Hermitian Berry Phase in Feedback-Coupled Oscillators}
\author{Yaashnaa Singhal}
\thanks{These authors contributed equally to this work.}
\affiliation{Department of Physics, University of Illinois at Urbana-Champaign, Urbana, IL 61801-3080, USA}
\author{Enrico Martello}
\thanks{These authors contributed equally to this work.}
\affiliation{School of Physics and Astronomy, University of Birmingham, Edgbaston, Birmingham B15 2TT, United Kingdom}
\author{Shraddha Agrawal}
\affiliation{Department of Physics, University of Illinois at Urbana-Champaign, Urbana, IL 61801-3080, USA}
\author{Tomoki Ozawa}
\email{tomoki.ozawa.d8@tohoku.ac.jp}
\affiliation{Advanced Institute for Materials Research (WPI-AIMR), Tohoku University, Sendai 980-8577, Japan}
\author{Hannah Price}
\email{H.Price.2@bham.ac.uk}
\affiliation{School of Physics and Astronomy, University of Birmingham, Edgbaston, Birmingham B15 2TT, United Kingdom}
\author{Bryce Gadway}
\email{bgadway@illinois.edu}
\affiliation{Department of Physics, University of Illinois at Urbana-Champaign, Urbana, IL 61801-3080, USA}
\date{\today}

\begin{abstract}
The geometrical Berry phase is key to understanding the behaviour of quantum states under cyclic adiabatic evolution. When generalised to non-Hermitian systems with gain and loss, the Berry phase can become complex, and should modify not only the phase but also the amplitude of the state. Here, we perform the first experimental measurements of the adiabatic non-Hermitian Berry phase, exploring a minimal two-site $\mathcal{PT}$-symmetric Hamiltonian that is inspired by the Hatano-Nelson model. We realise this non-Hermitian model experimentally by mapping its dynamics to that of a pair of classical oscillators coupled by real-time measurement-based feedback. As we verify experimentally, the adiabatic non-Hermitian Berry phase is a purely geometrical effect that leads to significant amplification and damping of the amplitude also for non-cyclical paths within the parameter space even when all eigenenergies are real.
We further observe a non-Hermitian analog of the Aharonov--Bohm solenoid effect, observing amplification and attenuation when encircling a region of broken $\mathcal{PT}$ symmetry that serves as a source of imaginary flux. This experiment demonstrates the importance of geometrical effects that are unique to non-Hermitian systems and paves the way towards the further studies of non-Hermitian and topological physics in synthetic metamaterials.
\end{abstract}
\maketitle

Geometrical phases play a fundamental role across physics as they
emerge from the cyclic adiabatic evolution of a system, and depend only on certain intrinsic geometrical properties within a given parameter space. In quantum mechanics, a key example of this is the Berry phase~\cite{berry1984quantal}, which can be related, not only to the quantum geometry of eigenstates, but also to important topological invariants, such as the Chern number and winding number~\cite{xiao2010,hasan2010}. Experimentally, the Berry phase has  profound effects on material and transport properties, and it underlies Hall effects, polarization, charge pumping, semiclassical dynamics and many other phenomena~\cite{xiao2010}.

Following its discovery, the Berry phase was generalised to systems with dissipation or gain, in which the Hamiltonian becomes non-Hermitian~\cite{garrison1988complex, dattoli1990geometrical,  Keck_2003, liang2013, mondragon1996berry, berry2003optical,berry2004physics, nesterov2008complex}. Interest in such problems has continued to grow, inspired by developments in non-Hermitian experimental platforms, including in
photonics~\cite{guo2009,Nori-NatMats-Review},
mechanics~\cite{Coulais2017,Brandenbourger2019,scheibner2020,scheibner2020odd,zhou2020non,Coulais-PNAS,MechMeta}, electric circuits~\cite{choi2018observation,helbig2020generalized}, and cold atoms~\cite{Bo-NonRecip,li2019observation} amongst many others~\cite{Coulais2021,ashida2020,bergholtz2021}. This progress has also been driven by interest in topological systems, in which non-Hermiticity leads to new topological classifications and unusual boundary phenomena~\cite{ Coulais2021, bergholtz2021}.

Underlying these effects are fundamental differences between Hermitian and non-Hermitian Hamiltonians; this includes that eigenstates can coalesce and become defective at exceptional points, that the left and right eigenfunctions will typically be different from each other, and that the eigenenergies can become complex~\cite{ashida2020,  Nori-NatMats-Review}. One important consequence of these differences is that the Berry phase will, in general, become complex- instead of real-valued, implying that the amplitude as well as the phase of a state will vary under adiabatic dynamical evolution~\cite{garrison1988complex,dattoli1990geometrical, Keck_2003,longhi2009bloch,liang2013, hayward2018, Ilan-BerryConnection,hayward2020}.

In this paper, we measure the adiabatic non-Hermitian  Berry phase, demonstrating how non-Hermiticity leads to gauge-invariant geometrical effects even for non-cyclical paths in parameter space. This goes beyond previous experiments which observed the real part of a Berry phase for closed loops around non-Hermitian exceptional points~\cite{dembowski2001,dembowski2004,gao2015observation}; in those cases, the Berry phase was parametric rather than adiabatic as
adiabaticity inevitably breaks down when an exceptional point is dynamically encircled~\cite{uzdin2011observability, berry2011optical,doppler2016dynamically} and geometrical properties have therefore to be reconstructed from eigenmode measurements. In contrast, here we study a two-site $\mathcal{PT}$-symmetric system, in a regime for which the eigenenergies are real and adiabatic evolution is possible. To realise our model, we employ a mapping between quantum evolution and the classical dynamics of a pair of oscillators coupled with real-time measurement-based feedback~\cite{MechMeta}. We evolve our system adiabatically and experimentally demonstrate that the imaginary part of the Berry phase leads to significant geometrical amplification and damping, which is intrinsically non-Hermitian.

\textit{Non-Hermitian Berry phase:}
Before discussing our experiment, we review the basic theory of non-Hermitian systems~\cite{ashida2020,  Nori-NatMats-Review} to motivate the non-Hermitian Berry phase. Various related definitions exist for this phase~\cite{garrison1988complex, dattoli1990geometrical,  Keck_2003, mondragon1996berry, berry2003optical,berry2004physics, nesterov2008complex,liang2013}; here, we introduce a formalism that is motivated by physical observables to concisely include all relevant geometrical effects using Berry connections. This definition has the advantage that its imaginary part is manifestly gauge-invariant and is immediately related to measurements of the population. Detailed derivations are given in the Supplemental Material.

We consider a $N$-component state vector $|\psi (t)\rangle$, which depends on time $t$ and obeys the Schr\"odinger-type equation $i\partial_t |\psi(t)\rangle = H(\boldsymbol\lambda) |\psi(t) \rangle$, where the family of $N$-by-$N$ non-Hermitian matrices $H(\boldsymbol\lambda)$ are parametrized by a set of real parameters $\boldsymbol\lambda = (\lambda_1, \lambda_2, \cdots)$. For a given value of $\boldsymbol\lambda$, $H(\boldsymbol\lambda)$ acts as a non-Hermitian Hamiltonian.
It has right and left eigenvectors, denoted by $|R_n (\boldsymbol\lambda)\rangle$ and $\langle L_n (\boldsymbol\lambda)|$ respectively, which are generally not complex conjugates of each other~\cite{ashida2020, Nori-NatMats-Review}, but which share the same complex eigenvalues $\varepsilon_n (\boldsymbol\lambda)$, indexed by $n = 1, 2, \cdots, N$.
Within the parameter space spanned by $\boldsymbol\lambda$, four distinct geometrical Berry connections can then be defined~\cite{shen2018,Ilan-BerryConnection};
however, for the non-Hermitian Berry phase, only the following two  Berry connections will be relevant:
\begin{align}
	\mathcal{A}^{LR}_{n,j}(\boldsymbol\lambda) &\equiv i\langle L_n (\boldsymbol\lambda)| \partial_{\lambda_j}|R_n (\boldsymbol\lambda)\rangle / \langle L_n (\boldsymbol\lambda)| R_n (\boldsymbol\lambda) \rangle,
	\\
	\mathcal{A}^{RR}_{n,j}(\boldsymbol\lambda) &\equiv i\langle R_n (\boldsymbol\lambda)| \partial_{\lambda_j}|R_n (\boldsymbol\lambda)\rangle / \langle R_n (\boldsymbol\lambda)| R_n (\boldsymbol\lambda) \rangle.
\end{align}
Upon a generalized gauge transformation, which multiplies $|R_n (\boldsymbol\lambda)\rangle$ and $|L_n (\boldsymbol\lambda)\rangle$ not just by a phase but also by arbitrary and independent nonzero factors, it can be shown that the following combination of the above Berry connections is invariant~\cite{Ilan-BerryConnection}:
\begin{align}
	\delta\mathcal{A}^{LR-RR}_{n,j}(\boldsymbol\lambda) &\equiv \mathcal{A}^{LR}_{n,j}(\boldsymbol\lambda) - \mathcal{A}^{RR}_{n,j}(\boldsymbol\lambda). \label{eq:gaugeinvcomb}
\end{align}
It is a distinguishing feature of non-Hermitian systems that gauge-independent quantities can be constructed just from a linear combination of Berry connections; in Hermitian quantum mechanics, the different Berry connections coincide and the gauge-invariant combination $\delta\mathcal{A}^{LR-RR}_{n,i}$ is always zero.

\begin{figure*}[t!]
	\includegraphics[width=1.9\columnwidth]{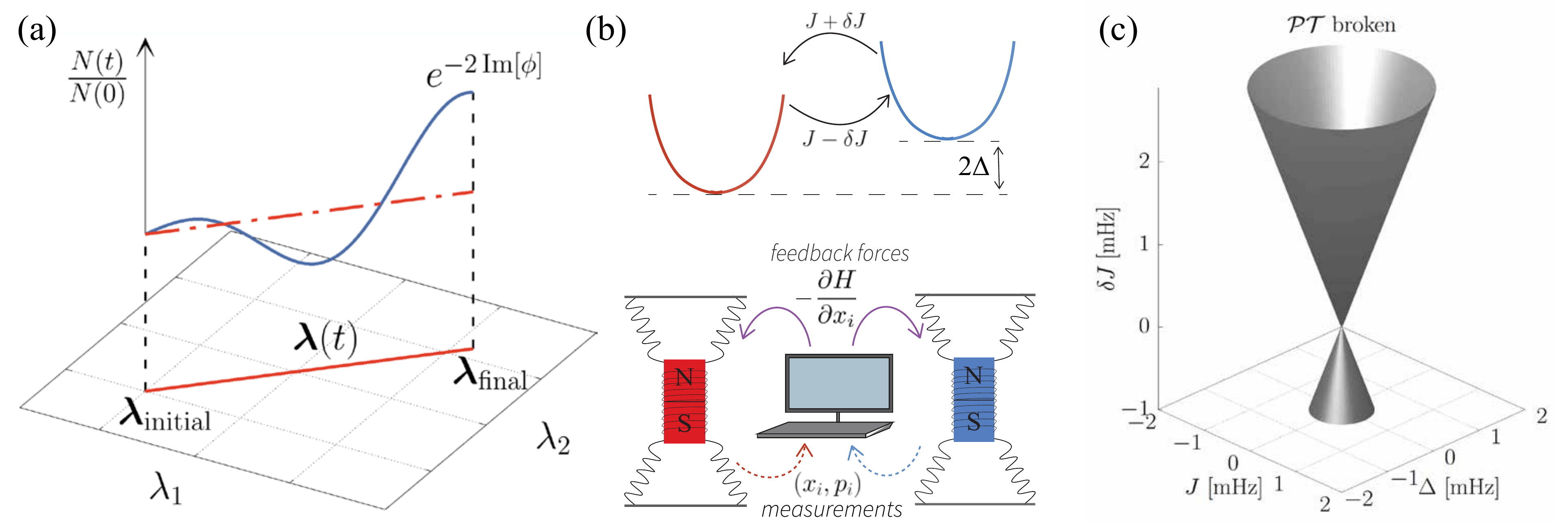}
	\centering
	\caption{\label{FIG:F1}
		\textbf{Exploration of the non-Hermitian Berry phase in the two-site Hatano-Nelson model.}
        \textbf{(a)}~In non-Hermitian systems with $\mathcal{PT}$ symmetry, adiabatic paths in parameter space generically results in amplification or attenuation of the time-dependent population $N (t)$. This results directly from the imaginary portion of the adiabatic non-Hermitian Berry phase $\phi$ acquired by a state along its trajectory.
        \textbf{(b)}~Top: The Hatano-Nelson (HN) dimer, a minimal non-Hermitian lattice model with non-reciprocal left/right hopping rates $J \pm \delta J$ and an inter-site frequency imbalance $2\Delta$.
        Bottom: Implementation of the HN dimer in a mechanical system via measurement-and-feedback. Non-reciprocal coupling between mechanical oscillators, as well as shifts to their resonance frequencies, are realised through applied forces that are responsive to real-time measurements.
        \textbf{(c)}~$\mathcal{PT}$ symmetry-breaking phase diagram of the HN dimer. A conical surface of exceptional points in the $J-\delta J-\Delta$ parameter space separates regions of broken and preserved $\mathcal{PT}$ symmetry, respectively lying inside and outside of the conical surface.
	}
\end{figure*}

We now consider the adiabatic evolution of a state upon changing the parameter $\boldsymbol\lambda (t)$ as a function of time $t$ to extract the non-Hermitian counterpart of the Berry phase~\cite{garrison1988complex,dattoli1990geometrical, massar1996,Keck_2003,longhi2009bloch,liang2013, hayward2018, Ilan-BerryConnection,hayward2020}.
Here, we focus on the situation where all the eigenvalues are real and non-degenerate so that we can apply the adiabatic theorem~\cite{nenciu1992adiabatic,hoeller2020non}. 
Then if the initial state corresponds to the $n$-th right eigenstate, the state at time $t$ can be written as
\begin{align}
	|\psi(t)\rangle = c(t)\frac{|R_n (\boldsymbol\lambda(t))\rangle}{\sqrt{\langle R_n (\boldsymbol\lambda(t))|R_n (\boldsymbol\lambda(t))\rangle}} \label{eq:psit}
\end{align}
where $c(t)$ is a complex-valued adiabatic factor that the state acquires as $\boldsymbol\lambda(t)$ is varied. In defining $c(t)$, we chose to separate out the denominator, as we are interested in physical observables such as the population $N(t)$, which is then given simply by $N(t)\equiv \langle \psi(t)|\psi(t)\rangle = |c(t)|^2$.
We note that the final result is independent of the way the state $|\psi (t)\rangle$ is written as a product of a coefficient $c(t)$ and a basis vector, as explained in detail in Supplemental Material.
We formally solve the Sch\"odinger equation $i\partial_t |\psi(t)\rangle = H(\boldsymbol\lambda(t))|\psi(t)\rangle = \varepsilon_n (\boldsymbol\lambda(t)) |\psi(t)\rangle$ by applying $\langle L_n(\boldsymbol\lambda(t))|$ from the left, 
which yields
\begin{align}
	c(t) = & \ c(0)\exp
	\left[
	-i\int_0^t dt^\prime \varepsilon_n(\boldsymbol\lambda(t^\prime))
	+ i\phi[\mathcal{C}]
	\right],
\end{align}
where the first term in the exponent is the dynamical contribution to the adiabatic factor $c(t)$, whereas the second part is the non-Hermitian Berry phase that we define by
\begin{align}
\phi [\mathcal{C}]
\equiv
\int_\mathcal{C}d\boldsymbol\lambda \cdot \left( \boldsymbol{\mathcal{A}}_n^{LR}(\boldsymbol\lambda) - i\mathrm{Im}\boldsymbol{\mathcal{A}}_n^{RR}(\boldsymbol\lambda) \right), \label{eq:berry}
\end{align}
where $\boldsymbol{\mathcal{A}}_n^{LR}(\boldsymbol\lambda) = (\mathcal{A}_{n,1}^{LR}(\boldsymbol\lambda), \mathcal{A}_{n,2}^{LR}(\boldsymbol\lambda), \cdots)$ and similarly for $\boldsymbol{\mathcal{A}}_n^{RR}(\boldsymbol\lambda)$.
The non-Hermitian Berry phase depends on the path $\mathcal{C}$ taken in parameter space and reflects the geometrical structure of the eigenstates, analogous to the well-known Berry phase for Hermitian systems~\cite{berry1984quantal,xiao2010}. However, unlike the Hermitian Berry phase, the non-Hermitian Berry phase has both real and imaginary parts. In particular, the imaginary part
\begin{align}
	\mathrm{Im}(\phi[\mathcal{C}]) &= \int_\mathcal{C}d\boldsymbol\lambda \cdot \mathrm{Im} \delta \boldsymbol{\mathcal{A}}_n^{LR-RR}(\boldsymbol\lambda),
\end{align}
depends solely on the imaginary part of $\delta \boldsymbol{\mathcal{A}}_n^{LR-RR}(\boldsymbol\lambda) = (\delta \mathcal{A}_{n,1}^{LR-RR}(\boldsymbol\lambda), \delta \mathcal{A}_{n,2}^{LR-RR}(\boldsymbol\lambda), \cdots)$, which is the gauge-invariant combination of Berry connections introduced in Eq.~(\ref{eq:gaugeinvcomb}).
Therefore, it is then immediately obvious that the imaginary part of the non-Hermitian Berry phase is gauge independent even when the path $\mathcal{C}$ is not closed~\cite{massar1996}. On the other hand, the real part of the Berry phase is gauge invariant only when the path $\mathcal{C}$ forms a closed path, just like in the Hermitian case~\cite{xiao2010}.
When the eigenvalues are all real, the evolution of the population [as depicted in Fig.~\ref{FIG:F1}~(a)] is thus determined purely by the imaginary part of the Berry phase as
\begin{align}
	N(t) = |c(t)|^2 = N(0)\exp \left[ -2\mathrm{Im}(\phi [\mathcal{C}])\right],
\end{align}
which is directly observable in our experiment.
%
%

\textit{Experimental set-up:}
To experimentally explore the effects of non-Hermitian geometry, we implement the simple two-site model Hamiltonian
\begin{align}
    H =
    \begin{pmatrix}
    -\Delta & J + \delta J \\ J - \delta J & \Delta
    \end{pmatrix} \ , \label{eq:model}
\end{align}
as depicted in Fig.~\ref{FIG:F1}~(b).
The elements of $H$ have units of frequency, consistent with the aforementioned Schr\"odinger-type equation describing the system dynamics.
Physically, the real parameters $\Delta$, $J$, and $\delta J$ relate to relevant frequency shifts of ($\Delta$) and hopping rates between ($J \pm \delta J$) the oscillators.
This model is inspired by the Hatano-Nelson model for a 1D lattice~\cite{HN-PRL-96}, which has non-reciprocal hoppings between neighbouring lattice sites and which can exhibit nontrivial topology and the non-Hermitian skin effect~\cite{bergholtz2021}. The eigenvalues of Eq.~\ref{eq:model} are given by $\varepsilon_\pm = \pm \sqrt{\Delta^2 + J^2 - \delta J^2}$, which means that the two eigenvalues are both real when $\Delta^2 + J^2 > \delta J^2$, corresponding to the $\mathcal{PT}$-symmetric region. If $\Delta^2 + J^2 = \delta J^2$, the eigenvalues coalesce at an exceptional point; within the parameter space of $(\Delta, J, \delta J)$, the surface of exceptional points corresponds to a double cone, with its apex at the origin~\cite{nesterov2008complex}, as shown in Fig.~\ref{FIG:F1}~(c).

The gauge-invariant combinations of the Berry connections within the $\mathcal{PT}$-symmetric region (c.f. Eq.~\ref{eq:gaugeinvcomb}) are all purely imaginary, and they diverge as we approach the $\mathcal{PT}$-symmetry breaking transition, where adiabaticity breaks down. (Analytical expressions of the Berry connections and associated Berry curvatures are derived in the Supplemental Material, and can be interpreted in terms of a complex hyperbolic pseudo-magnetic monopole in parameter space~\cite{nesterov2008complex,hayward2020}.) This in turn means that the only non-vanishing part of the non-Hermitian Berry phase (Eq.~\ref{eq:berry}) is purely imaginary and therefore gauge-invariant for any path.

To explore the two-site Hatano-Nelson model, we construct a synthetic mechanical lattice made up of two classical oscillators artificially coupled by real-time feedback measurements, based on our approach reported in Ref.~\cite{MechMeta}. The essential idea of this scheme is to map the Heisenberg equations for a desired tight-binding quantum Hamiltonian onto Newton’s equations of motion for classical oscillators in phase-space within a rotating wave-approximation~\cite{salerno2016,salerno2014dynamical}. As discussed in \cite{MechMeta}, the use of real-time feedback then means that many two level non-Hermitian Hamiltonian can be realized with this setup. Here, we use self- and cross-feedback between the oscillators to realize the Hamiltonian described in Eq.~\ref{eq:model}, as depicted at the bottom of Fig.~\ref{FIG:F1}~(b).
Self-feedback terms proportional to the oscillator positions ($F_i \propto x_i$) allow us to shift their frequencies by $\pm \Delta$ from a nominal starting value of $f_0 \approx 3.05$~Hz.
Cross-feedback forces ($F_i \propto x_{j}$) allow us to introduce independent left-to-right and right-to-left hopping terms $J \pm \delta J$, with no intrinsic limitation to reciprocal energy exchange.
By applying self-feedback terms proportional to the oscillator momenta ($F_i \propto p_i$), we cancel the oscillators' natural damping and explore coherent dynamics for well over 1000~s ($> 3000$ periods). These long timescales are crucial to performing the first explorations of adiabatic response in a non-Hermitian system.
Beyond single-body (quadratic, in the operator sense) terms, we additionally apply higher-order feedback to cancel nearly all native quartic nonlinearities. However, small residual nonlinearities remain, serving to, \textit{e.g.}, cap the energy growth in cases of broken $\mathcal{PT}$ symmetry.

\begin{figure}[t!]
	\includegraphics[width=0.98\columnwidth]{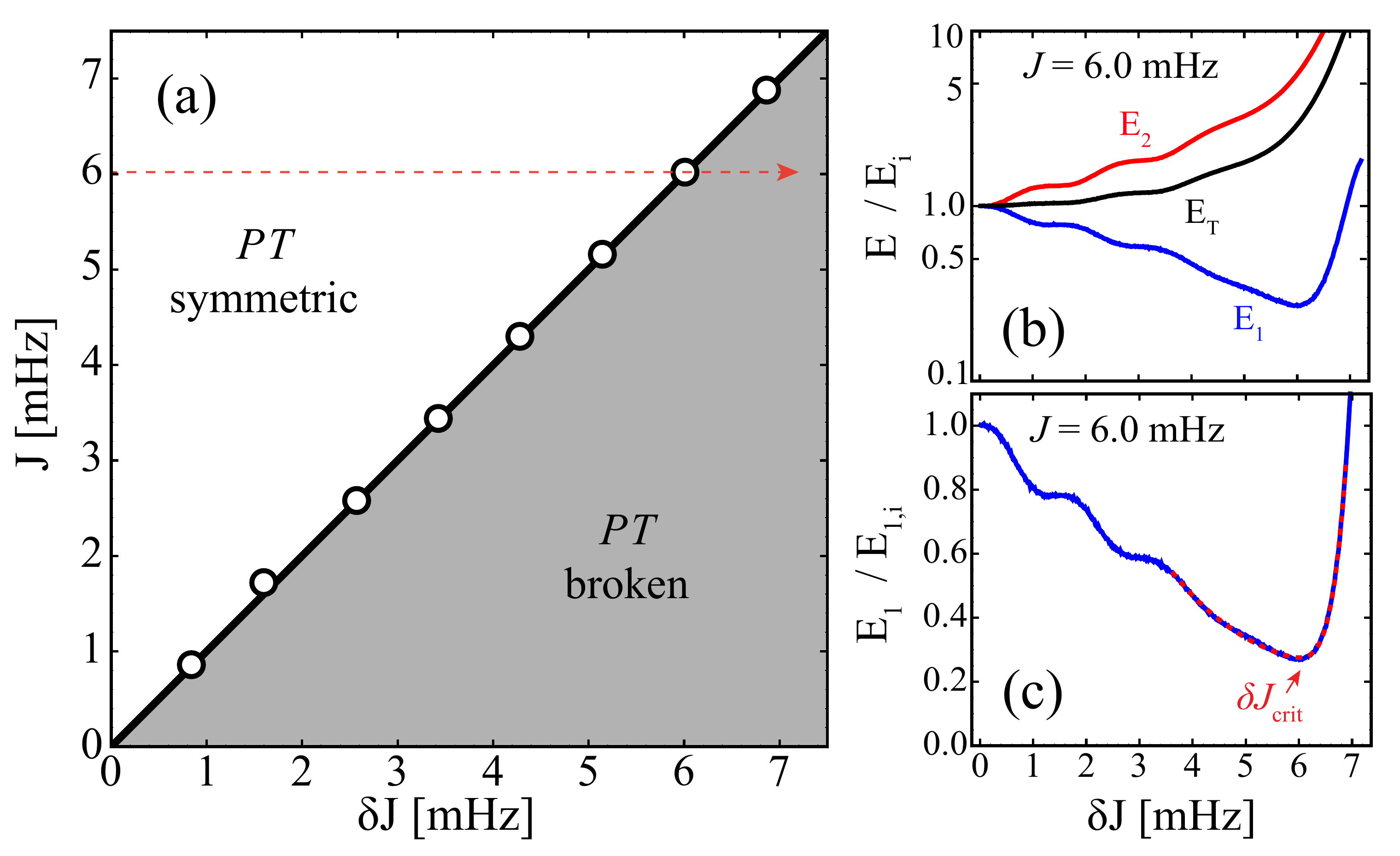}
	\centering
	\caption{\label{FIG:F2a}
		%
		\textbf{$\mathcal{PT}$ symmetry-breaking phase diagram of the unbiased ($\Delta = 0$) Hatano--Nelson dimer.}
        \textbf{(a)}~White points mark the experimentally measured exceptional points (EPs). Critical $\delta J$ values for these points are determined by detecting the breakdown of adiabaticity as the EP is crossed.
        \textbf{(b)}~Experimental energy dynamics of prepared eigenstates along the ramp of $\delta J$ [dashed red line in (a)] for $J = 6.0$~mHz. The measured energy at site 1 decays (while the site 2 and total energy grow) until the EP is reached at $\delta J \sim J$. Here, we plot the site 1 ($E_1$), site 2 ($E_2$), and total energy ($E_T$) normalized to their respective initial values ($E_i$).
        \textbf{(c)}~Crossing of the EP is marked by the onset of growth of the otherwise decaying site 1. The red dashed line is an empirical fit to the data, with the fit minimum defining $\delta J_\textrm{crit}$. Here, we plot the energy in oscillator 1 normalized to its initial value ($E_1/E_{1,i}$).
	}
\end{figure}

\begin{figure*}[t!]
	\includegraphics[width=1.85\columnwidth]{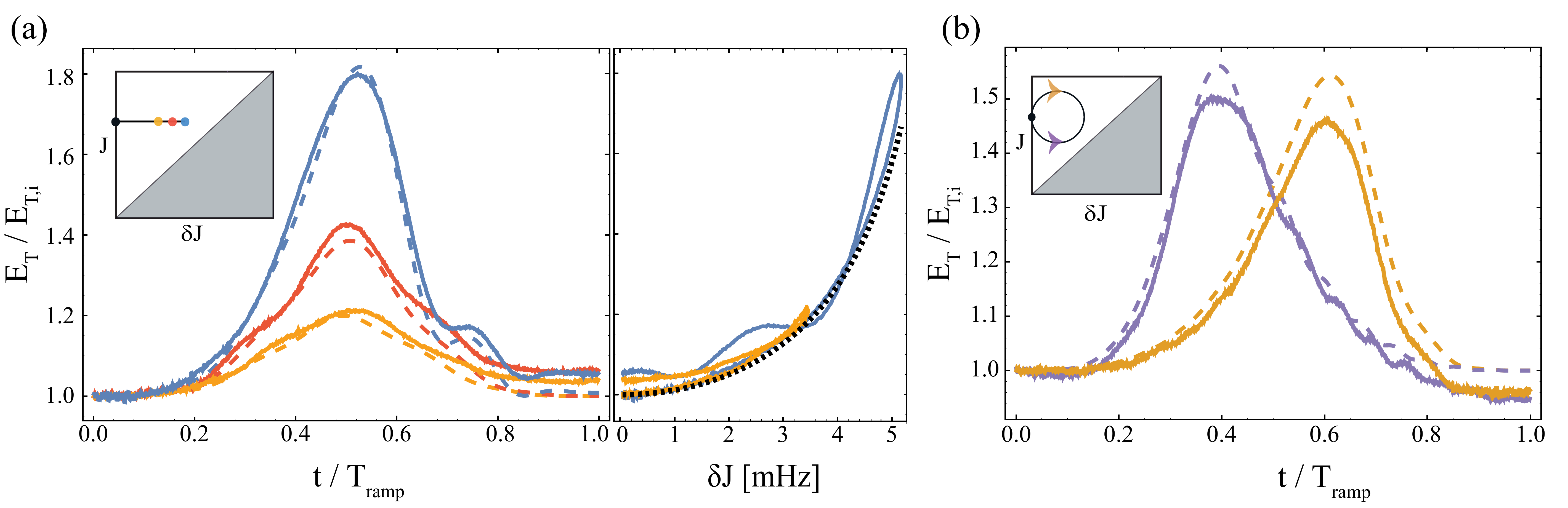}
	\centering
	\caption{\label{FIG:F2b}
		%
		\textbf{Geometric energy amplification and attenuation in the Hatano--Nelson dimer.}
		\textbf{(a)}~Left:~Dynamics of the total energy $E_T = E_1 + E_2$ for adiabatic transformations (normalized to its initial value $E_{T,i}$).
		Inset: path of the adiabatic transformation, for fixed $J = 6.5$~mHz, ramping from $\delta J = 0$ to/from different maximum values of $\delta J_\textrm{max}$~$=$~3.4~mHz~(yellow), $\delta J_\textrm{max}$~$=$~4.3~mHz~(red), and $\delta J_\textrm{max}$~$=$~5.2~mHz~(blue).
		Right:~Plot of the total energy vs. the instantaneous $\delta J$ value, for the yellow and blue paths. The black dotted line shows the expected parametric dependence of $E_T/E_{T,i}$ on $\delta J$ for fully adiabatic evolution under $H$.
		\textbf{(b)}~Dynamics of $E_T$ for CW (purple) and CCW (gold) paths as specified by the inset, centered at $(J,\delta J) = (6.5,2.2)$~mHz and having a radius of $2.2$~mHz.
		The colored dashed lines in the main panels relate to the time-dependent solutions of the Schr\"odinger-type equation based on evolution under the ideal HN model, including non-adiabatic effects due to the finite ramp durations of $T_\textrm{ramp} = 500~s$ for (a) and $T_\textrm{ramp} = 1000~s$ for (b).
	}
\end{figure*}

\textit{Results:} We first experimentally establish the $\mathcal{PT}$ symmetry-breaking phase diagram of the canonical two-site Hatano--Nelson model, with tunable reciprocal ($J$) and non-reciprocal ($\delta J$) components of the real-valued inter-site hopping, but with no inter-site bias ($\Delta = 0$). In this case, for fixed $J$, an exceptional point and $\mathcal{PT}$ symmetry-breaking phase transition are encountered at $\delta J = J$, as previously demonstrated with this platform by spectral analysis in Ref.~\cite{MechMeta}. In the full ($J$,$\delta J$) parameter space, there are two distinct regions of conserved and broken $\mathcal{PT}$ symmetry, denoted by white and grey in Fig.~\ref{FIG:F2a}~(a).
We experimentally determine the exceptional line separating these regions
by probing the breakdown of adiabaticity and the rapid onset of energy growth as states cross over into the $\mathcal{PT}$-broken region, as shown in Fig.~\ref{FIG:F2a}~(b).
We prepare eigenmodes of the symmetric double-well
for various fixed values of the reciprocal hopping $J$, and then linearly ramp $\delta J$ from 0 to 1.2~$J$ over 400~s.
We establish the exceptional points (white circles) by determining the instantaneous $\delta J$ values for which there begins to be energy growth at the otherwise decaying first site, as shown in Fig.~\ref{FIG:F2a}~(c).
Here, an observable proportional to the $i^\textrm{th}$ oscillator energy $E_i(t)$ is reconstructed from the measured $x_i$ and $p_i$ signals~\cite{MechMeta}.
We can then associate the oscillators' energy dynamics with relative changes in the macroscopic mechanical energy population $N_i(t) \sim E_i(t)/hf_0$.
We now restrict ourselves to the $\mathcal{PT}$ symmetric region of Fig.~\ref{FIG:F2a}~(a), exploring the adiabatic gauge invariant non-Hermitian Berry phase acquired (via the energy dynamics of prepared eigenmodes) as we slowly evolve along controlled paths in parameter space.
In Fig.~\ref{FIG:F2b}~(a), we
first prepare our system as an eigenmode of the symmetric double well for a fixed reciprocal hopping $J = 6.45$~mHz, and then
we smoothly vary the asymmetric hopping as $\delta J(t) = \delta J_\textrm{max}\sin^2 (\pi t / T_\textrm{ramp})$ over a time $T_\textrm{ramp} = 500$~s.
From the left inset, this corresponds to a closed linear path in parameter space from the black dot at $\delta J = 0$ to one of the colored dots (representing different values of $\delta J_\textrm{max}$), and back.
It is seen in Fig.~\ref{FIG:F2b}~(a) that the total energy
increases as the trajectory moves closer to the exceptional line, with the blue ($\delta J_\textrm{max} = 5.16$~mHz) path showing the largest gain. For such a trajectory, the energy in the system is determined by the instantaneous $\delta J$ value, as confirmed by the parametric collapse of the energy vs. $\delta J$ for the blue and yellow curves, shown in the right inset.
To note, slight wiggles in both the data (solid lines) and the numerical simulation curves (dashed lines, which include effects of the finite ramp duration) arise, primarily due to non-adiabatic deviations accumulated near the exceptional line.
However, for all curves the total energy returns to near its initial value at the end of the trajectories,
consistent with adiabatic evolution along a time-reversed path that encloses zero non-Hermitian flux.

In Fig.~\ref{FIG:F2b}~(b), we start from the same conditions but now move along closed circular loops by also varying the symmetric hopping term $J$ by a sinusoidal function over a time period of 1000~s. Coordination between the variation of $J$ and $\delta J$ allows us to make either clockwise or counterclockwise paths in parameter space (inset).
The energy dynamics curves for the two path directions are essentially (up to small non-adiabatic corrections) mirrored versions of each other with respect to the time midpoint $T_\textrm{ramp}/2$, as the gauge-invariant Berry phase accumulated from the common starting point is again uniquely determined by the instantaneous position in parameter space. This is consistent with the fact that these finite-area paths enclose \textit{zero} non-Hermitian Berry phase.
To note, the curves in Fig.~\ref{FIG:F2b}~(a) and Fig.~\ref{FIG:F2b}~(b) do exhibit percent level gain and loss over their respective evolution times of 500 and 1000~s, stemming from residual loss and gain terms at the scale of a few $\mu$Hz.
%

\begin{figure*}[t!]
	\includegraphics[width=1.95\columnwidth]{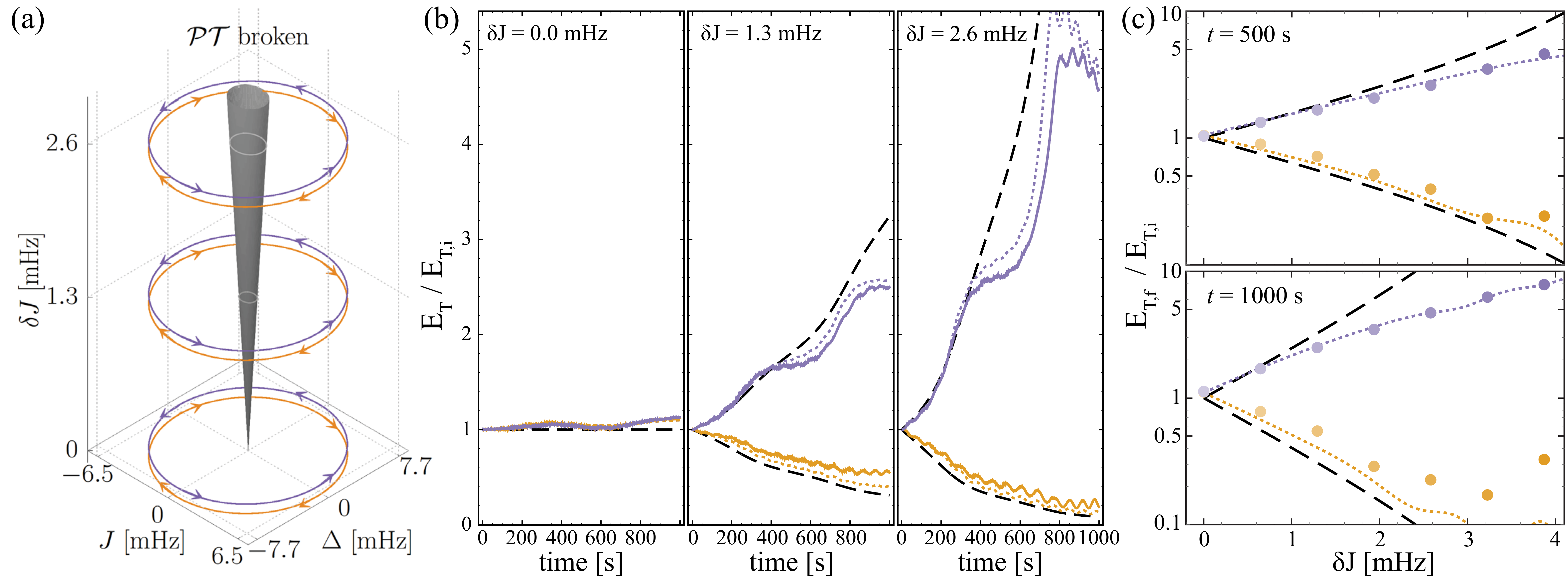}
	\centering
	\caption{\label{FIG:F3}
		\textbf{Cyclic amplification and attenuation along adiabatic paths enclosing $PT$-broken sources of non-Hermitian Berry phase.}
		\textbf{(a)}~Cartoon depiction of the paths traversed in parameter space. For trajectories in the $J-\Delta$ plane at fixed and finite $\delta J$, the paths enclose a conical $\mathcal{PT}$-broken region that acts as a source of non-Hermitian Berry flux.
		\textbf{(b)}~Dynamics of the total energy $E_T = E_1 + E_2$ for CCW (purple) and CW (gold) paths in the $J-\Delta$ plane as specified in panel (a), shown for three values of the tunneling asymmetry $\delta J$. The solid purple (gold) lines are the measured trajectories from experiment for CCW (CW) paths. The long-dashed and short-dashed lines are theory comparisons, described below.
		\textbf{(c)}~Ratio of the final total energy $E_{T,f}$ to the initial total energy $E_{T,i}$ as a function of the tunneling asymmetry $\delta J$. The upper and lower panels show the energy ratios for one half cycle and one full cycle, respectively. The purple (gold) points are the experimentally measured ratios for the CCW (CW) paths (with error bars smaller than the data points) and the long-dashed and short-dashed lines are theory comparisons.
		For (b,c), the black long-dashed lines are the analytical predictions (detailed in the Supplement) for the amplification/attenuation under fully adiabatic evolution according to Eq.~\ref{eq:model}. The dotted lines are the trajectories determined by numerical simulation of the experimental ramping procedure, also incorporating weak nonlinear contributions that serve to capture the saturation observed for large amplification (detailed in the Supplement).
	}
\end{figure*}

We now explore closed paths in parameter space that enclose a region of broken $\mathcal{PT}$ symmetry, and which correspondingly aquire a finite non-Hermitian Berry phase. We accomplish this by introducing a site-to-site energy bias ($\Delta$). To recall, the exceptional surface in the full $(\Delta, J, \delta J)$ parameter space of Eq.~\ref{eq:model} corresponds to a double cone with an apex at the origin~\cite{nesterov2008complex}. As depicted in Fig.~\ref{FIG:F3}~(a), this admits closed paths within the $\mathcal{PT}$ symmetric region that enclose areas of broken $\mathcal{PT}$ symmetry. The $\mathcal{PT}$-broken region can, in a sense, serve as a source of non-Hermitian Berry flux, in analogy to the manner in which a magnetic solenoid serves as a source of flux in
the canonical Aharonov--Bohm thought experiment~\cite{AB-effect}.
Indeed, our procedure can be viewed as measuring the imaginary Aharonov-Bohm phase in parameter space.

We explore the dynamics of the total energy $E_T$ as we traverse counter-clockwise (CCW) and clockwise (CW) paths in the $J-\Delta$ plane, starting from several fixed values of $\delta J$. We start by preparing eigenmodes of the system with $J = 0$ and then ramp, over 1000~s, about an ellipse in the $J-\Delta$ parameter space as displayed in Fig.~\ref{FIG:F3}~(a). As we see from Fig.~\ref{FIG:F3}~(b), the dynamics of the total energy are strongly dependent on $\delta J$. In the fully symmetric case, $\delta J = 0$, we find no significant change to the total oscillator energy, as expected from the lack of an enclosed $\mathcal{PT}$-broken region. For increasing values of $\delta J$, we find that the CCW (CW) paths in parameter space lead to an increasing growth (decay) of the energy upon completing one cycle. Figure~\ref{FIG:F3}~(c) summarizes the $\delta J$-dependence of the measured gain (attenuation) of the total energy experienced upon completing one cycle in the CCW (CW) direction.
The near-exponential dependence of the measured gain (attenuation) with $\delta J$ is in qualitative agreement with the expected variation of the acquired non-Hermitian Berry phase for cyclic paths.
The non-Hermitian Berry phase accumulated around such paths grows with the size of the $\mathcal{PT}$ broken region, having a form that is nearly proportional to $\delta J$, as presented in the Supplement.
For the experimentally traversed path in the CCW (CW) direction, the system picks up a negative (positive) contribution of this imaginary phase, and the state of the oscillators thus experiences a corresponding growth (decay) in its energy.
At short times or for small values of the hopping asymmetry $\delta J$, the observed amplification and attenuation are in fair agreement with the analytical form expected based on pure geometric contributions of an imaginary Berry phase. However, clear deviations can be found, most prominently in situations where very large growth of the total energy are expected (CCW orbits for large $\delta J$ values). On physical grounds, deviations from the expected response can be expected for very large oscillator displacements due to natural anharmonicities. We qualitatively capture the observed saturation of growth by comparing to a dynamical evolution that incorporates small but non-negligible (empirical) nonlinear contributions, described further in the Supplement.


\textit{Conclusion:} We have experimentally measured the non-Hermitian Berry phase for adiabatic evolution in a two-site Hatano-Nelson model. We have demonstrated significant geometrical contributions to amplification and damping along both closed and open paths, and shown that these effects are observable in a synthetic mechanical metamaterial. Going further, we will be able to add different types of nonlinearities to the two-site Hatano-Nelson model, allowing us to explore the interplay of interactions with $\mathcal{PT}$-symmetry~\cite{RMP-Konotop,Nonlinearity-Induced-PT-transition-Segev}. As active mechanical metamaterials are scaled up to larger systems with dozens of oscillators, they will enable controllable explorations of
the effects of quantum geometry and topology in non-Hermitian Chern insulators, such as, for example, the anomalous velocity contributions predicted to arise from the non-Hermitian Berry phase~\cite{Ilan-BerryConnection} and the breakdown of the canonical bulk-boundary correspondence of Hermitian models~\cite{Yao-NH-Chern,Kunst-NH-BulkBound,Kohei-NH-ChernIns}.

\section{Acknowledgements}
We thank Barry Bradlyn for helpful discussions.
This material (Y.~S., S.~A., B.~G.) is based upon work supported by the National Science Foundation under grant No.~1945031. Y.~S. acknowledges support by the Philip J. and Betty M. Anthony Undergraduate Research Award and the Jeremiah D. Sullivan Undergraduate Research Award of the UIUC Department of Physics. T.~O. acknowledges support from JSPS KAKENHI Grant No. JP20H01845, JST PRESTO Grant No. JPMJPR19L2, JST CREST Go. Number JPMJCR19T1, and RIKEN iTHEMS. E.~M. and H.~M.~P. are supported by the Royal Society via grants UF160112, RGF\textbackslash EA\textbackslash 180121 and RGF\textbackslash R1\textbackslash 180071.

%
%
%
%
%
%


%

\clearpage

\begin{widetext}
\begin{center}
\begin{large}
{\bf Supplemental Material for ``Measuring the Adiabatic Non-Hermitian Berry Phase in Feedback-Coupled Oscillators"}
\end{large}
\end{center}
\end{widetext}

\section{Theoretical background on non-Hermitian Berry phase}
We summarize here details of the theoretical formalism for the non-Hermitian Berry phase.

We consider a family of $N$-by-$N$ matrices $H(\boldsymbol\lambda)$ parametrized by a set of real parameters $\boldsymbol\lambda = (\lambda_1, \lambda_2, \cdots)$. For a given value of $\boldsymbol\lambda$, the matrix $H(\boldsymbol\lambda)$ is generally not Hermitian, and we consider a system which obeys the Schr\"odinger equation where $H(\boldsymbol\lambda)$ acts as a non-Hermitian Hamiltonian: $i\partial_t |\psi (t)\rangle = H(\boldsymbol\lambda)|\psi(t)\rangle$. Here, $|\psi (t)\rangle$ is a time-dependent $N$-component vector, which is an analog of the wavefunction in this non-Hermitian system.

Various properties of the system are characterized by the eigenvectors and eigenvalues of the non-Hermitian Hamiltonian $H(\boldsymbol\lambda)$. Assuming that the Hamiltonian is diagonalizable, there exist $N$ right eigenvectors $|R_n (\boldsymbol\lambda)\rangle$ and the same number of left eigenvectors $\langle L_n (\boldsymbol\lambda)|$ indexed by an integer $n = 1, 2, \cdots, N$, which satisfy $H(\boldsymbol\lambda)|R_n (\boldsymbol\lambda) \rangle = \varepsilon_n(\boldsymbol\lambda) |R_n(\boldsymbol\lambda)\rangle$ and $\langle L_n (\boldsymbol\lambda) | H(\boldsymbol\lambda) = \varepsilon_n(\boldsymbol\lambda) \langle L_n(\boldsymbol\lambda)|$. Here, generally complex eigenvalues $\varepsilon_n (\boldsymbol\lambda)$ are common for the right and left eigenvectors.
Assuming non-degeneracy of energies, namely $\varepsilon_n (\boldsymbol\lambda) \neq \epsilon_{n^\prime} (\boldsymbol\lambda)$ for $n \neq n^\prime$ for a given value of $\boldsymbol\lambda$, left and right eigenvectors of different eigenvalues are orthogonal $\langle L_n (\boldsymbol\lambda) | R_{n^\prime} (\boldsymbol\lambda)\rangle = 0$ if $n \neq n^\prime$. 
There is no {\it a priori} reason to take a particular normalization for the left and/or right eigenvectors, and thus, upon defining the eigenvectors, there is a freedom to choose an overall multiplicative factor.
Physically observable quantities should not depend on this choice of overall factors of the eigenvectors. In other words, we should look for properties which are independent under the following gauge transformations:
\begin{align}
	|R_n (\boldsymbol\lambda)\rangle &\to r_n (\boldsymbol\lambda) |R_n(\boldsymbol\lambda)\rangle,
	&
	| L_n (\boldsymbol\lambda) \rangle &\to l_n (\boldsymbol\lambda) | L_n (\boldsymbol\lambda)\rangle, \label{eq:gauge}
\end{align}
where $r_n (\boldsymbol\lambda)$ and $l_n (\boldsymbol\lambda)$ are nonzero complex numbers.

From the analogy with Hermitian quantum mechanics, we define the Berry connection for non-Hermitian systems; however, since the left and the right eigenvectors are different, we need to consider the following four distinct Berry connections~\cite{shen2018,Ilan-BerryConnection}:
\begin{align}
	\mathcal{A}^{LR}_{n,j}(\boldsymbol\lambda) &\equiv i\langle L_n (\boldsymbol\lambda)| \partial_{\lambda_j}|R_n (\boldsymbol\lambda)\rangle / \langle L_n (\boldsymbol\lambda)| R_n (\boldsymbol\lambda) \rangle,
	\\
	\mathcal{A}^{RL}_{n,j}(\boldsymbol\lambda) &\equiv i\langle R_n (\boldsymbol\lambda)| \partial_{\lambda_j}|L_n (\boldsymbol\lambda)\rangle / \langle R_n (\boldsymbol\lambda)| L_n (\boldsymbol\lambda) \rangle,
	\\
	\mathcal{A}^{RR}_{n,j}(\boldsymbol\lambda) &\equiv i\langle R_n (\boldsymbol\lambda)| \partial_{\lambda_j}|R_n (\boldsymbol\lambda)\rangle / \langle R_n (\boldsymbol\lambda)| R_n (\boldsymbol\lambda) \rangle,
	\\
	\mathcal{A}^{LL}_{n,j}(\boldsymbol\lambda) &\equiv i\langle L_n (\boldsymbol\lambda)| \partial_{\lambda_j}|L_n (\boldsymbol\lambda)\rangle / \langle L_n (\boldsymbol\lambda)| L_n (\boldsymbol\lambda) \rangle.
\end{align}
Upon the gauge transformation Eq.~(\ref{eq:gauge}), these four Berry connections transform as
\begin{align}
	\mathcal{A}^{LR}_{n,j}(\boldsymbol\lambda) &\to \mathcal{A}^{LR}_{n,j}(\boldsymbol\lambda) + i\partial_{\lambda_j} \ln r_n (\boldsymbol\lambda) \label{eq:berryconnection1}
	\\
	\mathcal{A}^{RL}_{n,j}(\boldsymbol\lambda) &\to \mathcal{A}^{RL}_{n,j}(\boldsymbol\lambda) + i\partial_{\lambda_j} \ln l_n (\boldsymbol\lambda)
	\\
	\mathcal{A}^{RR}_{n,j}(\boldsymbol\lambda) &\to \mathcal{A}^{RR}_{n,j}(\boldsymbol\lambda) + i\partial_{\lambda_j} \ln r_n (\boldsymbol\lambda)
	\\
	\mathcal{A}^{LL}_{n,j}(\boldsymbol\lambda) &\to \mathcal{A}^{LL}_{n,j}(\boldsymbol\lambda) + i\partial_{\lambda_j} \ln l_n (\boldsymbol\lambda). \label{eq:berryconnection4}
\end{align}
Thus, each Berry connection is gauge-dependent, namely the Berry connections depend on how one chooses the normalization of the left and right eigenvectors. However, we can construct the following gauge-invariant combinations of the Berry connections~\cite{Ilan-BerryConnection}:
\begin{align}
	\delta\mathcal{A}^{LR-RR}_{n,j}(\boldsymbol\lambda) &\equiv \mathcal{A}^{LR}_{n,j}(\boldsymbol\lambda) - \mathcal{A}^{RR}_{n,j}(\boldsymbol\lambda) \label{eq:gaugeinvcomb1}
	\\
	\delta\mathcal{A}^{RL-LL}_{n,j}(\boldsymbol\lambda) &\equiv \mathcal{A}^{RL}_{n,j}(\boldsymbol\lambda) - \mathcal{A}^{LL}_{n,j}(\boldsymbol\lambda).
\end{align}
We note that it is a distinguishing feature of non-Hermitian systems that we can construct gauge-independent quantities just from the Berry connections. Since the four Berry connections coincide for Hermitian quantum mechanics, the gauge-invariant combinations $\delta\mathcal{A}^{LR-RR}_{n,j}$ and $\delta\mathcal{A}^{RL-LL}_{n,j   }$ are zero when the Hamiltonian is Hermitian. As we see soon, these gauge-independent combinations of Berry connections appear in a properly-defined Berry phase upon adiabatic change of the parameter $\boldsymbol\lambda$.

We now consider an adiabatic evolution of a state upon changing the parameter $\boldsymbol\lambda (t)$ as a function of time $t$ and study a scaling factor that the state acquires, which is a non-Hermitian counterpart of the Berry phase~\cite{garrison1988complex,dattoli1990geometrical, massar1996,Keck_2003,longhi2009bloch,liang2013, hayward2018, Ilan-BerryConnection,hayward2020}.  In Hermitian systems, the adiabatic theorem guarantees that, if the change of the parameters is sufficiently slow, the state remains in the original state up to an overall phase factor if the initial state is an energy eigenstate. As mentioned above, one needs to be cautious when applying the adiabatic theorem to non-Hermitian systems; however, we now assume that this theorem can be applied. Starting from a right eigenstate, we consider adiabatic change of the parameter $\boldsymbol\lambda(t)$ as the time $t$ changes, and follow how the eigenstate evolves during this process. We take a family of right eigenstates $|R_n (\boldsymbol\lambda(t))\rangle$. Assuming that the adiabatic theorem holds, if the initial state is in the $n$-th right eigenstate, the state at time $t$ can be written as
\begin{align}
	|\psi(t)\rangle = c(t)\frac{|R_n (\boldsymbol\lambda(t))\rangle}{\sqrt{\langle R_n (\boldsymbol\lambda(t))|R_n (\boldsymbol\lambda(t))\rangle}} \label{eq:psiadi}
\end{align}
where $c(t)$ is a complex function accounting for the adiabatic factor the state acquires as
$\boldsymbol\lambda(t)$ is varied.
The denominator is introduced so that the change of the intensity of the state $|\psi (t)\rangle$ can be captured by just looking at the coefficient $c(t)$ because $\langle \psi(t)|\psi(t)\rangle = |c(t)|^2 \equiv N(t)$, where $N(t)$ is the population.
In the next section, we show that the final result is independent of how $|\psi (t)\rangle$ is written as a product of the coefficient and a basis vector.

We can formally solve the Schr\"odinger equation $i\partial_t |\psi(t)\rangle = H(\boldsymbol\lambda(t))|\psi(t)\rangle = \varepsilon_n (\boldsymbol\lambda(t)) |\psi(t)\rangle$ to obtain the Berry phase. We first apply $\langle L_n(\boldsymbol\lambda(t))|$ from the left to the Schr\"odinger equation. Writing the state as Eq.(\ref{eq:psiadi}), the Schr\"odinger equation becomes
\begin{align}
	&i\langle L_n (\boldsymbol\lambda(t))| \frac{\partial}{\partial t}
	\left(
	c(t)\frac{|R_n (\boldsymbol\lambda(t))\rangle}{\sqrt{\langle R_n (\boldsymbol\lambda(t))|R_n (\boldsymbol\lambda(t))\rangle}}
	\right)
	\notag \\
	&=
	c(t) \varepsilon_n (\boldsymbol\lambda(t))
	\frac{\langle L_n (\boldsymbol\lambda(t))|R_n (\boldsymbol\lambda(t))\rangle}{\sqrt{\langle R_n (\boldsymbol\lambda(t))|R_n (\boldsymbol\lambda(t))\rangle}}.
\end{align}
Expanding the time derivative and arranging terms, we obtain the following differential equation for $c(t)$:
\begin{align}
	\frac{\partial c(t)}{\partial t}
	=
	&c(t)
	\left[
	-i\varepsilon_n (\boldsymbol\lambda(t))
	- \frac{\langle L_n (\boldsymbol\lambda(t))|\partial_t | R_n (\boldsymbol\lambda(t))\rangle}{\langle L_n (\boldsymbol\lambda(t))|R_n (\boldsymbol\lambda(t))\rangle}
	\right.
	\notag \\
	&\left.
	+\frac{1}{2}
	\frac{\langle R_n (\boldsymbol\lambda(t))|\partial_t | R_n (\boldsymbol\lambda(t))\rangle + c.c.}{\langle R_n (\boldsymbol\lambda(t))|R_n (\boldsymbol\lambda(t))\rangle}
	\right],
\end{align}
where $c.c.$ stands for complex conjugate of the preceding term.
Converting the time derivative in the right-hand side to the derivative in $\boldsymbol\lambda$, and using the definition of the Berry connections, Eq.(\ref{eq:berryconnection1})-(\ref{eq:berryconnection4}), we obtain
\begin{align}
\frac{\partial c(t)}{\partial t}
=
&c(t)
\left[
-i\varepsilon_n (\boldsymbol\lambda(t)) + \phantom{\frac{1}{2}} \right. \notag \\
&\left.
+ i\frac{\partial \boldsymbol\lambda (t)}{\partial t}\cdot
\left( \boldsymbol{\mathcal{A}}^{LR}_n (\boldsymbol\lambda) -i \mathrm{Im} \boldsymbol{\mathcal{A}}^{RR}_n (\boldsymbol\lambda) \right)
\right],
\end{align}
where $\boldsymbol{\mathcal{A}}_n^{LR}(\boldsymbol\lambda) = (\mathcal{A}_{n,1}^{LR}(\boldsymbol\lambda), \mathcal{A}_{n,2}^{LR}(\boldsymbol\lambda), \cdots)$ and similarly for $\boldsymbol{\mathcal{A}}_n^{RR}(\boldsymbol\lambda)$.
Solving this differential equation, we obtain
\begin{align}
	c(t) =& c(0)\exp
	\left[
	-i\int_0^t dt^\prime \varepsilon_n(\boldsymbol\lambda(t^\prime))
	+
	\right.
	\notag \\
	&\left.
	i\int_\mathcal{C}d\boldsymbol\lambda \cdot \left( \boldsymbol{\mathcal{A}}_n^{LR}(\boldsymbol\lambda) -i \mathrm{Im}\boldsymbol{\mathcal{A}}_n^{RR}(\boldsymbol\lambda) \right)
	\right],
\end{align}
The first term in the exponent is the dynamical contribution to the adiabatic factor $c(t)$, whereas the second part is the line integral in the parameter space and only depends on the path $\mathcal{C}$,
and is thus analogous to the Berry phase, reflecting the geometrical structure of the eigenstates. In analogy with Hermitian quantum mechanics, we define the non-Hermitian Berry phase as a function of the path $\mathcal{C}$
as
\begin{align}
	\phi [\mathcal{C}]
	=
	\int_\mathcal{C}d\boldsymbol\lambda \cdot \left( \boldsymbol{\mathcal{A}}_n^{LR}(\boldsymbol\lambda) - i\mathrm{Im}\boldsymbol{\mathcal{A}}_n^{RR}(\boldsymbol\lambda) \right).
\end{align}
Unlike the Hermitian Berry phase, the non-Hermitian Berry phase has both the real and imaginary parts
\begin{align}
	\mathrm{Re}\left( \phi[\mathcal{C}] \right) &=
	\int_\mathcal{C}d\boldsymbol\lambda \cdot \mathrm{Re} \boldsymbol{\mathcal{A}}_n^{LR}(\boldsymbol\lambda),
	\\
	\mathrm{Im}\left( \phi[\mathcal{C}] \right) &= \int_\mathcal{C}d\boldsymbol\lambda \cdot \mathrm{Im} \delta \boldsymbol{\mathcal{A}}_n^{LR-RR}(\boldsymbol\lambda),
\end{align}
where $\delta \boldsymbol{\mathcal{A}}_n^{LR-RR}(\boldsymbol\lambda) = (\delta \mathcal{A}_{n,1}^{LR-RR}(\boldsymbol\lambda), \delta \mathcal{A}_{n,2}^{LR-RR}(\boldsymbol\lambda), \cdots)$ is the gauge-invariant combination of the Berry connections introduced in Eq.~(\ref{eq:gaugeinvcomb1}).
Therefore, the imaginary part of the non-Hermitian Berry phase is gauge independent even when the path $\mathcal{C}$ is not closed~\cite{massar1996}. On the other hand, the real part of the Berry phase is gauge invariant only when the path $\mathcal{C}$ forms a closed path, just like in the Hermitian case.

An advantage of the above formulation in terms of the Berry connection is also that the geometrical contribution to the time evolution of the population $N(t) = |c(t)|^2$ is manifestly gauge-invariant even for an open path:
\begin{align}
	N(t) = N(0)\exp \left[ 2\int_0^t dt^\prime \mathrm{Im} \varepsilon_n (\boldsymbol\lambda (t^\prime)) - 2\mathrm{Im}\left( \phi[\mathcal{C}] \right) \right]. \label{eq:nt}
\end{align}
Therefore, when the energy eigenvalue is real during the time evolution and hence $\mathrm{Im} \varepsilon_n (\boldsymbol\lambda (t)) = 0$, which is the case we experimentally study in this paper, the population dynamics is solely determined by the imaginary part of the Berry phase. And thus by observing the population dynamics, we can experimentally determine the imaginary part of the Berry phase.

\section{Different definition of the adiabatic coefficient}
The observable result of the non-Hermitian Berry phase through the population dynamics, Eq.~(\ref{eq:nt}), is insensitive to the way the adiabatic coefficient $c(t)$ is defined. For demonstration, we write, instead of as in Eq.~(\ref{eq:psiadi}), the state vector as
\begin{align}
|\psi(t)\rangle = \tilde{c}(t) |R_n (\boldsymbol\lambda (t))\rangle,
\end{align}
which is also a natural definition to take.
Applying $\langle L_n (\boldsymbol\lambda(t))|$ from left to the Schr\"odinger equation
\begin{align}
    i\frac{\partial}{\partial t}\left\{ \tilde{c}(t) |R_n (\boldsymbol\lambda (t))\rangle \right\}
    =
    \varepsilon_n (\boldsymbol\lambda (t)) \tilde{c}(t) |R_n (\boldsymbol\lambda (t))\rangle,
\end{align}
we obtain the differential equation for $\tilde{c}(t)$:
\begin{align}
    \frac{\partial \tilde{c}(t)}{\partial t}
    =
    \tilde{c}(t) \left[ -i\varepsilon_n (\boldsymbol\lambda (t)) - \frac{\langle L_n (\boldsymbol\lambda(t))|\partial_t | R_n (\boldsymbol\lambda(t))\rangle}{\langle L_n (\boldsymbol\lambda(t))|R_n (\boldsymbol\lambda(t))\rangle} \right].
\end{align}
Integrating this differential equation, we obtain
\begin{align}
    \tilde{c}(t)
    =
    \tilde{c}(0)
    \exp
    \left[
    -i \int_0^t \varepsilon_n (\boldsymbol\lambda (t^\prime))dt^\prime
    + i \int_{\mathcal{C}}\boldsymbol{\mathcal{A}}_{n}^{LR}(\boldsymbol\lambda) \cdot d\boldsymbol\lambda
    \right]. \label{eq:ctildet}
\end{align}
On the other hand, the population evolves in time as
\begin{align}
    N(t) = |\tilde{c}(t)|^2 \langle R_n (\boldsymbol\lambda(t))| R_n (\boldsymbol\lambda(t)) \rangle.
\end{align}
To understand the dynamics, in addition to the time evolution of $\tilde{c}(t)$, we need to understand how $\langle R_n (\boldsymbol\lambda(t))| R_n (\boldsymbol\lambda(t)) \rangle$ evolves in time. By a simple calculation, we can show
\begin{align}
    &\frac{\partial}{\partial t}\left\{ \langle R_n (\boldsymbol\lambda(t))| R_n (\boldsymbol\lambda(t)) \rangle \right\}
    \notag \\
    &=
    \langle R_n (\boldsymbol\lambda(t))| R_n (\boldsymbol\lambda(t)) \rangle
    2\frac{\partial \boldsymbol\lambda (t)}{\partial t}
    \mathrm{Im}\left[ \boldsymbol{\mathcal{A}}_n^{RR}(\boldsymbol\lambda)\right].
\end{align}
Integrating this differential equation, we obtain
\begin{align}
    &\langle R_n (\boldsymbol\lambda(t))| R_n (\boldsymbol\lambda(t)) \rangle
    \notag \\
    &=
    \langle R_n (\boldsymbol\lambda(0))| R_n (\boldsymbol\lambda(0)) \rangle
    \exp \left[2 \int_{\mathcal{C}} \mathrm{Im}\left[ \boldsymbol{\mathcal{A}}_n^{RR}(\boldsymbol\lambda)\right]\cdot d \boldsymbol\lambda \right].
    \label{eq:rt}
\end{align}
Combining Eq.(\ref{eq:ctildet}) and Eq.(\ref{eq:rt}), time evolution of the population is
\begin{align}
    N(t) = N(0)
    &\exp\left[2\int_0^t \mathrm{Im}\varepsilon_n (\boldsymbol\lambda (t^\prime)) dt^\prime\right.\notag \\
    &\left.
    -2\int_{\mathcal{C}}\mathrm{Im} \left[ \boldsymbol{\mathcal{A}}_n^{LR}(\boldsymbol\lambda) - \boldsymbol{\mathcal{A}}_n^{RR}(\boldsymbol\lambda)\right]\cdot d\boldsymbol\lambda
    \right],
\end{align}
which is exactly the same as Eq.~(\ref{eq:nt}). Thus, the time evolution of the population is independent of the way the adiabatic coefficient is defined, as expected.

\section{Two-site Hatano-Nelson model}

Here we describe the two-site non-Hermitian model we experimentally realize. The Hamiltonian is
\begin{align}
    H/h =
    \begin{pmatrix}
    -\Delta & J + \delta J \\ J - \delta J & \Delta
    \end{pmatrix}. \label{eq:modelsup}
\end{align}
Here, we take $\Delta$, $J$ and $\delta J$ to be real parameters, relating to relevant frequency shifts of and hopping rates between the oscillators. The eigenvalues of Eq.~\ref{eq:modelsup} are given by $\varepsilon_\pm = \pm \sqrt{\Delta^2 + J^2 - \delta J^2}$, which means that the two eigenvalues are both real when $\Delta^2 + J^2 > \delta J^2$, corresponding to the $\mathcal{PT}$-symmetric region. If $\Delta^2 + J^2 = \delta J^2$, the eigenvalues coaelesce at an exceptional point; within the parameter space of $\boldsymbol\lambda = (\Delta, J, \delta J)$, this exceptional point corresponds to a double cone, with its apex at the origin~\cite{nesterov2008complex}, as shown in Fig.~\ref{FIG:F1}~(c).

Within the $\mathcal{PT}$-symmetric region, the gauge-invariant combinations of the Berry connections are
\begin{align}
    \delta \mathcal{A}_{\pm,\Delta}^{LR-RR}
    &=
    i\frac{\delta J}{2}\cdot \frac{\Delta \delta J \pm J\sqrt{\Delta^2 + J^2 - \delta J^2}}{(\Delta^2 + J^2)(\Delta^2 + J^2 - \delta J^2)} \label{eq:adelta}
    \\
    \delta \mathcal{A}_{\pm,J}^{LR-RR}
    &=
    i\frac{\delta J}{2}\cdot \frac{J\delta J \mp \Delta \sqrt{\Delta^2 + J^2 - \delta J^2}}{(\Delta^2 + J^2)(\Delta^2 + J^2 - \delta J^2)} \label{eq:aj}
    \\
    \delta \mathcal{A}_{\pm,\delta J}^{LR-RR}
    &=
    -i\frac{\delta J}{2}\cdot \frac{1}{\Delta^2 + J^2 - \delta J^2}, \label{eq:dj}
\end{align}
We note that they are all purely imaginary, and that they diverge as we approach the $\mathcal{PT}$-symmetry breaking transition,
where adiabaticity breaks down. The associated geometrical Berry curvature $\delta \boldsymbol {\Omega}_{\pm}^{LR-RR} \equiv \nabla \times  \delta \boldsymbol{\mathcal{A}}_{\pm}^{LR-RR}$, where $\delta \boldsymbol{\mathcal{A}}_{\pm}^{LR-RR} = (\delta \mathcal{A}_{\pm,\Delta}^{LR-RR}, \delta \mathcal{A}_{\pm,J}^{LR-RR}, \delta \mathcal{A}_{\pm,\delta J}^{LR-RR})$ is also imaginary and takes the form~\cite{nesterov2008complex}:
\begin{equation}
\delta \boldsymbol{\Omega}_{\pm}^{LR-RR} = \frac{\pm i}{2(\Delta^2 + J^2 - \delta J^2)^{3/2}}(\Delta, J, \delta J), \label{eq:curvature}
\end{equation}
which is analogous to the magnetic field of a one-sheeted complex hyperbolic magnetic monopole. From the above quantities, we are able to analytically calculate the non-Hermitian Berry phase and to predict the consequent amplification and damping of the population.

\section{Non-Hermitian Berry phase for paths enclosing $PT$-broken regime}
We provide theoretical analysis of the non-Hermitian Berry phase for the configuration described in Fig.~\ref{FIG:F3} of the main text. The path we take is an ellipse in $\Delta - J$ plane with a fixed value of $\delta J$. For a given value of $\delta J$, the path can be parametrized by a single parameter $\theta$ as
\begin{align}
    \Delta &= \Delta_0 \cos \theta,
    &
    J &= J_0 \sin \theta, \label{eq:ellipticpath}
\end{align}
where, in the experiment, $\theta$ depends on time $t$, which varies from 0 to $T$. Time dependence of $\theta(t)$ is such that initially, $\theta(0) = 0$ and at the final time, $\theta (T) = \Theta \le 2\pi$. Note that the path is not closed unless $\Theta = 2\pi$. In our experiment, $\Delta_0 = 7.5$mHz and $J_0 = 6.45$mHz.
We note that the gauge-invariant combinations of Berry connections $\delta \mathcal{A}_{\pm, \Delta}^{LR-RR}$ and $\delta \mathcal{A}_{\pm, J}^{LR-RR}$ are both purely imaginary in $PT$-symmetric region. Thus the imaginary part of the Berry phase along the path $\mathcal{C}$ described by Eq.(\ref{eq:ellipticpath}) is
\begin{align}
    &\mathrm{Im}[\phi[\mathcal{C}]]
    =
    -i\int_\mathcal{C} d\boldsymbol\lambda \cdot \delta\boldsymbol{\mathcal{A}}_{\pm}^{LR-RR}(\boldsymbol\lambda)
    \notag \\
    &=
    -i\int_0^{\Theta} d\theta
    \left\{ \partial_{\theta}\Delta(\theta) \delta\mathcal{A}_{\pm,\Delta}^{LR-RR}(\theta)
    +
    \partial_{\theta}J(\theta) \delta\mathcal{A}_{\pm,J}^{LR-RR}(\theta)
    \right\}
    \notag \\
    &=
    -i\int_0^{\Theta} d\theta
    \left\{ -\Delta_0 \sin \theta \delta\mathcal{A}_{\pm,\Delta}^{LR-RR}(\theta)
    \right.
    \notag \\
    &\left.
    \hspace{3cm}+
    J_0 \cos \theta \delta \mathcal{A}_{\pm,J}^{LR-RR}(\theta)
    \right\}.
\end{align}
Using expressions (\ref{eq:adelta}) and (\ref{eq:aj}), we can perform the integral and obtain
\begin{align}
    &\mathrm{Im}[\phi[\mathcal{C}]]
    =
    -\frac{1}{4}\log
    \left\{
    \frac{(\Delta_0^2 - \delta J^2)(\delta_0^2 \cos^2 \Theta + J_0^2 \sin^2 \Theta)}{\Delta_0^2 (\Delta_0^2 \cos^2 \Theta + J_0^2 \sin^2 \Theta - \delta J^2)}
    \right\}
    \notag \\
    &
    \mp \frac{\delta J}{2}\frac{J_0}{\Delta_0\sqrt{\Delta_0^2 - \delta J^2}}
    \Pi \left( 1 - \frac{J_0^2}{\Delta_0^2};\Theta \right| \left. 1 - \frac{J_0^2 - \delta J^2}{\Delta_0^2 - \delta J^2} \right), \label{eq:phic}
\end{align}
where $\Pi (n;\phi | k)$ is the incomplete elliptic integral of the third kind defined by
\begin{align}
    \Pi (n;\phi | m)
    =
    \int_0^\phi \frac{d\theta}{(1-n\sin^2 \theta)\sqrt{1 - m \sin^2 \theta}}.
\end{align}

For a closed path with $\Theta = 2\pi$, the real part of the non-Hermitian Berry phase also becomes gauge invariant (up to $2\pi$ multiples of an integer). However, we can show that for the path we consider here, the real part is zero, and thus the non-Hermitian Berry phase for a closed path in our setup is purely imaginary.
Furthermore, the first logarithm term in Eq.(\ref{eq:phic}) vanishes for a closed path and the elliptic integral becomes the complete one, yielding
\begin{align}
    &\phi[\mathcal{C}]_{\Theta = 2\pi}
    =
    \notag \\ &
    \mp \frac{i 2 \delta J J_0}{\Delta_0\sqrt{\Delta_0^2 - \delta J^2}}
    \Pi \left( 1 - \frac{J_0^2}{\Delta_0^2} \right| \left. 1 - \frac{J_0^2 - \delta J^2}{\Delta_0^2 - \delta J^2} \right), \label{eq:phicclosed}
\end{align}
where we now have the complete elliptic integral of the third kind defined by
\begin{align}
    \Pi (n | m)
    =
    \int_0^{\pi/2} \frac{d\theta}{(1-n\sin^2 \theta)\sqrt{1 - m \sin^2 \theta}}.
\end{align}
The expressions (\ref{eq:phic}) and (\ref{eq:phicclosed}) are what we used in Fig.~\ref{FIG:F3} of the main text to draw theoretical curves.

\begin{figure}[t!]
	\includegraphics[width=0.8\columnwidth]{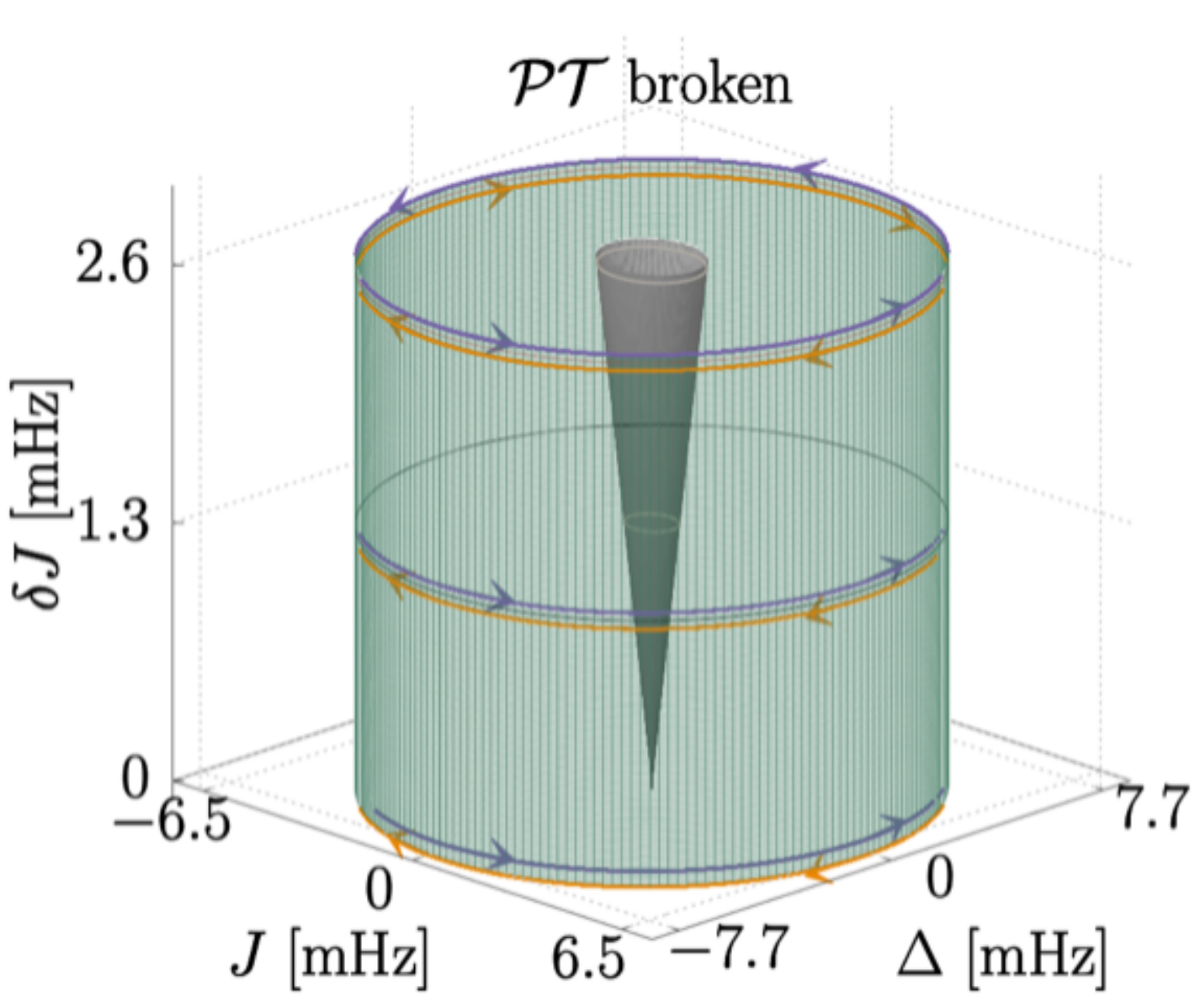}
	\centering
	\caption{\label{FIG:cylinder}
	The cylindrical surface of integral over which the Berry curvature can be defined everywhere continuous. The integral of the Berry curvature over this surface yields the non-Hermitian Berry phase corresponding to the closed elliptical path at a fixed value of $\delta J$.
	}
\end{figure}

So far, we have derived the non-Hermitian Berry phase through the line integral of the Berry connections along the path. For a closed path ($\Theta = 2\pi$) we can obtain the same result using the surface integral of the Berry curvatures as we show below. We can convert the line integral to the surface integral using Stokes' theorem, which is a standard procedure also in Hermitian cases. However, we should be careful that, in the plane determined by a fixed value of $\delta J$ in the parameter space of $\boldsymbol\lambda = (\Delta, J, \delta J)$, the interior of the curve $\mathcal{C}$ contains the $PT$-broken region. At the $PT$-breaking transition the eigenstates coalesce and we are no longer able to use the Stokes' theorem, which assumes the involved functions be continuous. We can, however, deform the surface of integral to the cylindrical form, as illustrated in Fig.~\ref{FIG:cylinder}. The surface of the cylinder is entirely contained in the $PT$-symmetric region, and thus Stokes' theorem can be safely applied.
The surface to be integrated contains two parts: side and the base.
The Berry curvature (\ref{eq:curvature}) does not have any component perpendicular to the base of the cylinder at $\delta J = 0$. We then need to consider the surface integral only over the side of the cylinder. The side of the cylinder can be parametrized by $\boldsymbol\lambda = (\Delta, J, \delta J) = (\Delta_0 \cos \theta, J_0 \sin \theta, \delta J^\prime)$ with parameters $\theta$ and $\delta J^\prime$. Here, $\theta$ varies from 0 to $2\pi$, and $\delta J^\prime$ varies from $\delta J$, which is the initial value, to $0$. Then, denoting the unit vector normal to the surface by $\mathbf{n}$ and the surface element by $dS$, the surface integral is
\begin{widetext}
\begin{align}
	\phi[\mathcal{C}]_{\Theta = 2\pi}
	&=
	\int \int \delta \boldsymbol\Omega_{\pm}^{LR-RR} \cdot \mathbf{n}dS
	\notag \\
	&=
	\int_0^{2\pi}d\theta \int_{\delta J}^0 d \delta J^\prime \delta \boldsymbol\Omega_{\pm}^{LR-RR} \cdot \left( \frac{\partial \boldsymbol\lambda}{\partial \theta} \times \frac{\partial \boldsymbol\lambda}{\partial \delta J^\prime} \right) d\theta d\delta J^\prime
	\notag \\
	&=
	\int_0^{2\pi}d\theta \int_{\delta J}^0 d \delta J^\prime
	\frac{\pm i \Delta_0 J_0}{2(\Delta_0^2 \cos^2 \theta + J_0^2 \sin^2 \theta - \delta J^{\prime 2})^{3/2}}
	\notag \\
	&=
	\int_0^{2\pi}d\theta
	\frac{\mp i \Delta_0 J_0 \delta J}{2(\Delta_0^2 \cos^2 \theta + J_0^2 \sin^2 \theta)\sqrt{\Delta_0^2 \cos^2 \theta + J_0^2 \sin^2 \theta - \delta J^{2}}}
	\notag \\
	&=
	\mp i \frac{J_0}{\Delta_0}\frac{2 \delta J}{\sqrt{\Delta_0^2 - \delta J^2}}
	\int_0^{\pi/2}d\theta
	\frac{1}{\left( 1- \left(1- \frac{J_0^2}{\Delta_0^2}\right) \sin^2 \theta \right)\sqrt{1 - \frac{\Delta_0^2 - J_0^2}{\Delta_0^2 - \delta J^2} \sin^2 \theta}}
	\notag \\
	&=
	\mp i \frac{J_0}{\Delta_0}\frac{2 \delta J}{\sqrt{\Delta_0^2 - \delta J^2}} \Pi\left( 1 - \frac{J_0^2}{\Delta_0^2} \right|\left. \frac{\Delta_0^2 - J_0^2}{\Delta_0^2 - \delta J^2} \right),
\end{align}
\end{widetext}
which is exactly the expression Eq.(\ref{eq:phicclosed}).
The population increase/decrease due to the imaginary part of the non-Hermitian Berry phase is thus directly related to the imaginary magnetic flux enclosed in the surface defined by the interior of the closed path. The change of population upon adiabatically following a closed path is a manifestation of the Aharonov-Bohm effect with an imaginary magnetic flux.

\section{Data acquisition and analysis}
As described previously in Ref.~\cite{MechMeta}, we acquire real-time proxies of the oscillators' positions based on continuous voltage signals from analog accelerometers attached to each oscillator. We additionally acquire a proxy for the oscillators' momenta based on determining the instantaneous jerk by taking the numerical derivative of acquired acceleration signals. These voltage signals proportional to the acceleration and jerk are normalized to a common scale, and respectively serve as dimensionless proxies for the oscillator positions ($\tilde{x_i}$) and momenta ($\tilde{p_i}$).
From the rapidly oscillating (at frequency $f_0$) $\tilde{x_i}$ and $\tilde{p_i}$ signals, we can reconstruct a proxy for the mechanical energy stored in each oscillator $E_i \propto \tilde{x}_i^2 + \tilde{p}_i^2$. Furthermore, the total energy of the two-oscillator dimer can be reconstructed simply as $E_{T} = E_1 + E_2$.

\begin{figure*}[t!]
	\includegraphics[width=1.9\columnwidth]{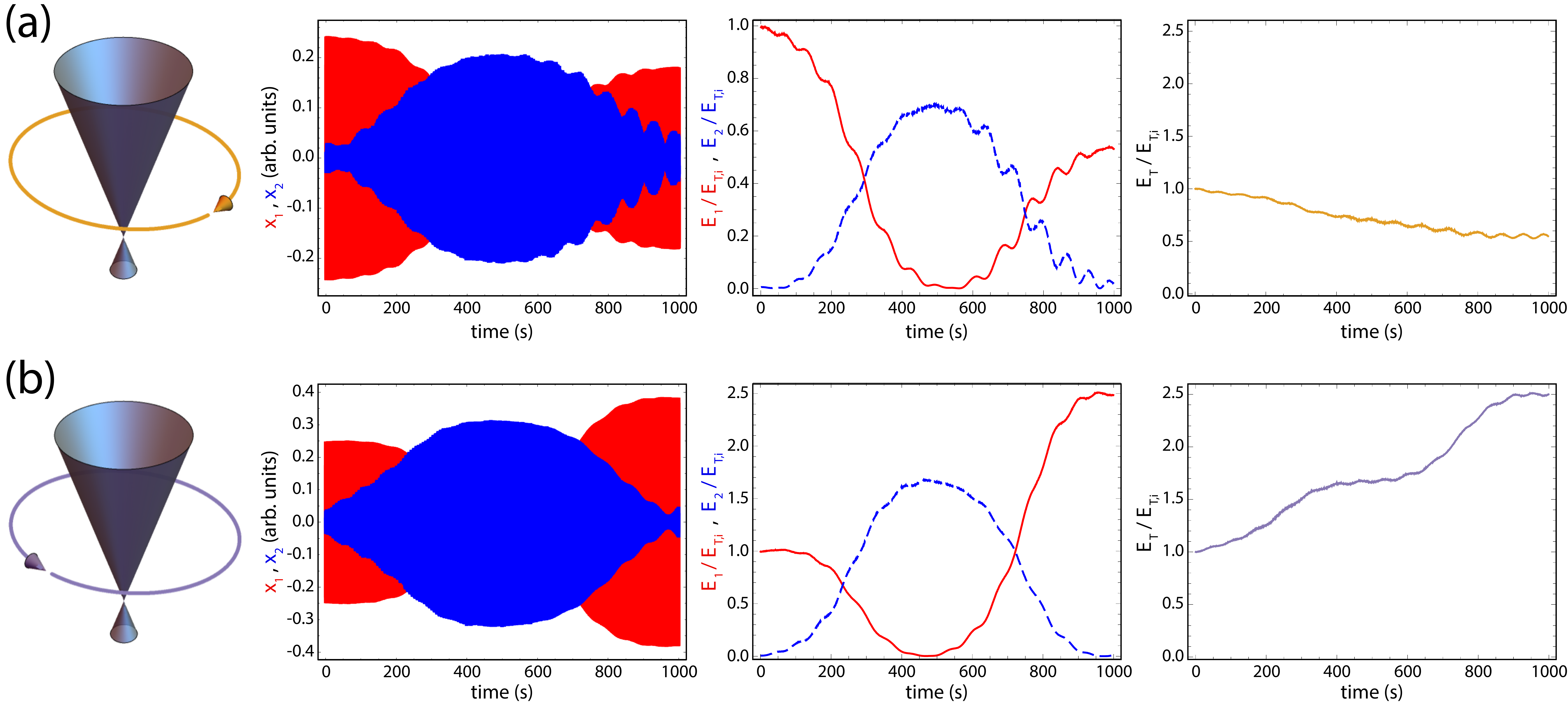}
	\centering
	\caption{\label{FIG:S1}
		\textbf{Representative primary and processed data, demonstrating how quantities like total energy are derived from the primary acquired signals for the individual oscillators.}
        Shown for the data as presented in Fig.~\ref{FIG:F3}, for $\delta J = 1.3$~mHz and for both directions of encircling.
        \textbf{Row (a):}~For CW (gold) trajectories in the $J-\Delta$ plane, the data panels display (from left to right) the progression from position ($\tilde{x}_i$) signals, to the individual energy signals in each oscillator, to the total energy of the two-oscillator dimer. To note, both the individual and total oscillator energies are normalized to the total energy at time $t=0$ (start of the encircling process). For the CW trajectory, the total energy is attenuated upon completing one cycle.
        \textbf{Row (b):}~Same data progression for the case of CCW (purple) trajectories in the $J-\Delta$ plane. For the CCW trajectory, the total energy is amplified upon completing one cycle.
	}
\end{figure*}

In Fig.~\ref{FIG:S1}, we display the progression of acquired and processed data, relating to the scenario of cyclic attenuation (a) and amplification (b) of energy upon encircling a $PT$-broken region. In both cases, we start with a state initialized by resonantly depositing energy into oscillator 1 while maintaining an inter-oscillator bias (frequency mismatch) $2\Delta$. The first column depicts the paths for CW (gold, row (a)) and CCW  (purple, row (b)) trajectories in the model's parameter space. The second column depicts the $\tilde{x}_i$ position signals for oscillators 1 (red) and 2 (blue). The $\tilde{p}_i$ momentum signals, which are indistinguishable from the $\tilde{x}_i$ signals on this long timescale, are not shown. The third column displays the calculated energies for each oscillator, normalized to the initial total energy. The fourth and final column displays the total mechanical energy in the two-oscillator dimer, normalized to the initial total energy. To note, the data from the final column is the same as that appearing (combined) in the central panel of Fig.~\ref{FIG:F3}~(b).

\section{Influence of weak nonlinearities}
Our experimental implementation involves rather large, few cm-scale displacements from equilibrium of our masses on springs (further details in Ref.~\cite{MechMeta}). For large displacements, there is a natural increase of the oscillation frequency due to the anharmonicity of the physical springs that appear above and below our oscillating masses. We apply feedback forces to cancel out, to first order, this natural stiffening effect. Specifically, for each oscillator, we apply a nonlinear contribution to the feedback forces of the form $F^{NL}_i = \alpha_i (\tilde{x}_i^2 + \tilde{p}_i^2)\tilde{x}_i$. In general, the control over $\alpha_i$ for such a nonlinear feedback can be used to implement nonlinear interaction terms, as was demonstrated in Ref.~\cite{MechMeta}. Here, we set $\alpha_i$ for each oscillator so as to cancel their natural quartic nonlinearities.

However, this cancellation is imperfect, and residual and higher-order nonlinearities lead to shifts in the frequencies of the oscillators that become more severe as the mechanical energy in an oscillator grows. In particular, for experiments in which we expect to observe a large increase in the overall mechanical energy due to the accrual of a complex Berry phase, we tend to find less growth than expected. Specifically, for the experiments described in Fig.~\ref{FIG:F3} of the main text involving cyclic amplification along CCW paths encircling the $PT$-broken region, we find reasonable agreement between our experimental data and the analytical theory results in cases in which the total energy remains small ($E_T / E_{T,i} \lesssim 2$, with $E_{T,i}$ being the initial total energy). However, for conditions in which there is larger energy growth, we observe significant deviations from the expected analytical results. To capture this effect, we also compare our data (for Fig.~\ref{FIG:F3}) to numerical simulation curves that incorporate additional nonlinear contributions to the dynamics. Specifically, we compare to dynamical evolution according to the set of coupled equations:
\begin{widetext}
\begin{align}
\dot{a} = -i [(\omega_0 - \Delta) a + i \Gamma_a a - (J + \delta J) b + \tilde{g} (|a|^2 - 1) a]  \\
\dot{b} = -i [(\omega_0 + \Delta) b + i \Gamma_b b - (J - \delta J) a + \tilde{g} (|b|^2 - 1) b]
\end{align}
\end{widetext}
The majority of terms are as defined previously in the text, with $\omega_0 = 2\pi f_0$ describing the common bare angular frequency of the two oscillators, $2\Delta$ the angular frequency mismatch imposed between the oscillators, and $J \pm \delta J$ the left-to-right and right-to-left hopping rates. These parameters are evolved according to the state preparation procedure and parameter ramps as described in the context of Fig.~\ref{FIG:F3}. Specifically, population is first initialized at site $a$ (corresponding to $a(t=0) = 1$) in the presence of a large initial detuning of $2\Delta_0$, with $\Delta_0/2\pi = 7.7$~mHz. Then, a purely non-reciprocal hopping of $\delta J$ is smoothly turned on as $\delta J = (\delta J_\textrm{max}/2)\times(1-\cos(\pi t/T_\textrm{prep}))$ over a preparation time of $T_\textrm{prep} = 250$~s, with variable final values of $\delta J_\textrm{max}$ as described in the main text. Then, after this preparation stage, encircling orbits in the $J-\Delta$ plane are made as depicted in Fig.~\ref{FIG:F3}. Specifically, the coordinates $\Delta$ and $J$ follow their respective forms of $\Delta = \Delta_0 \cos(2\pi t / T_\textrm{ramp})$ and $J = C \times J_0 \sin(2\pi t / T_\textrm{ramp})$, where $T_\textrm{ramp} = 1000$~s, $J_0/2\pi = 6.45$~mHz, and $C = \pm 1$ defines the chirality for CW and CCW parameter trajectories.

The fixed terms $\Gamma_{a,b}$ account for small residual loss/gain at oscillators $a$ and $b$, taking values of $\Gamma_{a}/2\pi$ = 10.25~$\mu$Hz and $\Gamma_{a}/2\pi$ = -5.75~$\mu$Hz, relating to weak loss and gain terms, respectively. The coefficient of the empirical nonlinear term contains both a real and imaginary contribution (\textit{i.e.}, $\tilde{g} = \tilde{g}_\textrm{re} + i \tilde{g}_\textrm{im}$), which take respective values of $\tilde{g}_\textrm{re}/2\pi = 1.12$~mHz and $\tilde{g}_\textrm{im}/2\pi = 32.5$~$\mu$Hz. To note, these values ($\Gamma_{a,b}$ and $\tilde{g}$) were determined based on a global optimization of the agreement between theory and experiment (minimization of summed squared residuals) for the full CCW ramp case for all $\delta J$ values.

The complex numbers $a$ and $b$ represent the field coefficients for the two mechanical oscillator modes. By convention these coefficients are normalized as $|a(t)|^2 + |b(t)|^2 = 1$ at time $t=0$, and as such their squared norms can be compared to the individual normalized (to the total initial energy) oscillator energies as presented in the main text ($E_1 / E_\textrm{T,i}$ and $E_2 / E_\textrm{T,i}$). As the frequencies and damping/gain of the individual oscillators were separately fine-tuned in experiment under the respective conditions of $|a(t=0)|^2 = 1$ and $|b(t=0)|^2 = 1$, the contributions from the empirical nonlinear terms vanish under those conditions. Importantly, the nonlinear contributions to the dynamical evolution become significant when the total energy of either oscillator grows large, reproducing the saturation of cyclic growth as observed in the main text.
%
%
%
\end{document}